\documentclass[showpacs,twocolumn,pre]{revtex4}
\usepackage[dvips]{epsfig}

\newcommand{\be}{\begin{equation}}
\newcommand{\ee}{\end{equation}}

\newcommand{\clL}{{\cal L}}
\newcommand{\hH}{\hat{H}}
\newcommand{\bfx}{{\bf x}}

\newcommand{\bea}{\begin{eqnarray}}
\newcommand{\eea}{\end{eqnarray}}
\newcommand{\hh}{\tilde{h}}
\newcommand{\prt}{\partial}

\newcommand{\rgl}{\rangle}
\newcommand{\lgl}{\langle}

\begin{document}

\title{On fractional time quantum dynamics}

\author{Alexander Iomin }

\affiliation{Department of Physics, Technion, Haifa, 32000,
Israel}

\date{\today}
\begin{abstract}
Application of the fractional calculus to quantum processes is
presented. In particular, the quantum dynamics is considered in
the framework of the fractional time Schr\"odinger equation (SE),
which differs from the standard SE by the fractional time
derivative: $\frac{\prt}{\prt t}\rightarrow
\frac{\prt^{\alpha}}{\prt t^{\alpha}}$. It is shown that for
$\alpha =1/2$ the fractional SE is isospectral to a comb model. An
analytical expression for the Green functions of the systems are
obtained. The semiclassical limit is discussed.

\end{abstract}


\maketitle

Application of the fractional calculus to quantum processes is a
new and fast developing part of quantum physics which studies
nonlocal quantum phenomena
\cite{kusnezov,laskin1,WBG,laskin2,hermann,levy,west,naber,tarasov,iomin}.
It aims to explore non-local effects found for either long-range
interactions or time-dependent processes with many scales
\cite{WBG,hilfer,bouchaud,klafter,zas1,zas2,sokolov}. Fractional
calculus with a variety of applications
\cite{WBG,hilfer,klafter,zas1,zas2,sokolov,mainardi,podlubny,oldham},
including applications to quantum processes
\cite{laskin1,laskin2,west,naber,tarasov,iomin,DongXu,WangXu,bhatti,lenzi},
is a well developed and well established field that was
extensively reviewed.

The concept of differentiation of non-integer orders rises from
works of Leibniz, Liouville, Riemann, Grunwald and Letnikov, see
\textit{e.g.}, \cite{podlubny,oldham}. Its application is related
to random processes with power law distributions. This corresponds
to the absence of characteristic average values for processes
exhibiting many scales \cite{klafter,shlesinger}. A continuously
increasing list of applications in many sciences has developed: it
includes material science \cite{hilfer,zt8}, physical kinetics
\cite{zas1}, anomalous transport theory with a variety of
applications in solid state physics
\cite{bouchaud,klafter,shlesinger} and in nonlinear dynamics
\cite{zas1,zas2}.

In quantum physics, the fractional concept can be introduced by
means of the Feynman propagator for non-relativistic quantum
mechanics as for Brownian path integrals \cite{feynman}.
Equivalence between the Wiener and the Feynman path integrals,
established by Kac \cite{kac}, indicates some relation between the
classical diffusion equation and the Schr\"odinger equation.
Therefore, an appearance of the space fractional derivatives in
the Schr\"odinger equation is natural, since both the standard
Schr\"odinger equation and the space fractional one obey the
Markov process. As shown in the seminal papers \cite{laskin1,
west}, it relates with the path integrals approach.  As a result
of this, the path integral approach for L\'evy stable processes,
leading to the fractional diffusion equation, can be extended to a
quantum Feynman-L\'evy measure which leads to the space fractional
Schr\"odinger equation \cite{laskin1,west}.

A fractional time derivative can be introduced in the quantum
mechanics by analogy with the fractional Fokker-Planck equation
(FFPE), as well, by means of the Wick rotation of time
$t\rightarrow -it/\hbar$ \cite{naber}. However its physical
interpretation is still vague: for example, a phase of the wave
function as well as the semiclassical approximation should be
understood. The fractional time Schr\"odinger equation was first 
considered in \cite{naber}. Its generalization to space-time 
fractional quantum dynamics \cite{DongXu,WangXu} was performed  and a 
relation to the fractional uncertainty \cite{bhatti} was studied as 
well. Exact solutions were also obtained for the time fractional 
nonlinear Schrodinger equation \cite{OMA}. It is worth noting that, 
contrary
to the space fractional derivative, the fractional time Schr\"odinger
equation describes non-Markovian evolution with a memory effect.

The fractional time quantum dynamics with the Hamiltonian
$\hat{H}(x)$ is described by the fractional Schr\"odinger equation
(FSE)
\be\label{fse1} %
(i\hh)^{\alpha}\frac{\prt^{\alpha}\psi(\bfx,t)}{\prt
t^{\alpha}}= \hat{H}\psi(\bfx,t)\, , %
\ee %
where $\alpha\leq 1$. For concordance of the dimension in Eq.
(\ref{fse1}) all variables and parameters are considered
dimensionless, and $\hh$ is the dimensionless Planck constant, see
also \cite{naber,DongXu}. For $\alpha=1$, Eq. (\ref{fse1}) is the
"conventional" (standard) Schr\"odinger equation. For $\alpha<1$
the fractional derivative is a formal notation of an integral with
a power law memory kernel of the form \cite{add1}
\be\label{fse2} \frac{\prt^{\alpha}\psi(t)}{\prt t^{\alpha}}\equiv
I_t^{1-\alpha}\frac{\prt\psi(t)}{\prt t}
=\int_0^t\frac{(t-\tau)^{-\alpha}}{\Gamma(1-\alpha)}
\frac{\prt\psi(\tau)}{\prt\tau}d\tau\, , %
\ee %
which is known as the Caputo fractional derivative
\cite{mainardi}, and $\Gamma(z)$ is a gamma function. This
definition makes it possible to carry out the Laplace transform of
the fractional derivative. Introducing the Laplace image
$\tilde{\psi}(s)=\hat{\clL}\psi(t)$, one obtains
\be\label{fse2_1} %
\hat{\clL}\Big[\frac{\prt^{\alpha}\psi(t)}{\prt t^{\alpha}}\Big]=
s^{\alpha}\tilde{\psi}(s)-s^{\alpha-1}\psi(0)\, . %
\ee %
Another interesting property of the FSE is time evolution in the
form of the Mittag-Leffler function. For the time-independent
Hamiltonian, the eigenvalue equation (with corresponding boundary
conditions) is $H\phi_{\lambda}=\lambda\phi_{\lambda}$. Therefore
one obtains the Green function in the term of the Mittag-Leffler
function $E_{\alpha,1}(z)\equiv E_{\alpha}(z)
=\sum_{j=0}^{\infty}z^j/\Gamma(j\alpha+1)$ \cite{klafter,naber}:
\be\label{fse3}
G(\bfx,t;\bfx')=\sum_{\lambda}
\phi_{\lambda}^*(\bfx')\phi_{\lambda}(\bfx)E_{\alpha}
\left(\lambda\left[\frac{t}{i\hh}\right]^{\alpha}\right)\, , %
\ee %
which is a fractional generalization of Green's function. It is
worth noting that this solution for the Green function in the form
of the Mittag-Leffler function does not satisfy Stone's theorem
on one-parameter unitary groups \cite{stone}.

In the general case, when the eigenvalue problem cannot be solved
rigorously, the analysis of the fractional Green function meets
serious deficiencies. For a example, the semiclassical analysis of
the FSE leads to a much more complicated form for the wave
function than the physically transparent expression for the local
wave function $\psi(x,t)\sim e^{iS(t,x)/\hbar}$. Moreover, the
classical action $S(t,x)$ is not defined anymore for the FSE.
Another important question could be addressed to both FSE
(\ref{fse1}) and Eq. (\ref{fse3}): for the fractional
Fokker-Planck equation, the fractional time derivative is an
asymptotic description of a continuous time random walk and it
describes subdiffusion. Therefore, what is the physical meaning of
the fractional time derivative in quantum processes when a concept
of a multi-scale continuous time random walk is absent?

To shed light on this situation, we consider a case with
$\alpha=1/2$, when fractional quantum dynamics can be modelled by
means of the conventional quantum mechanics in the framework of a
comb model. The comb model is an analogue of a 1d medium where
fractional diffusion has been observed \cite{em1,bi2004}. It is a
particular example of a non-Markovian phenomenon, explained in the
framework of a so-called continuous time random walk
\cite{em1,shlesinger,klafter}. This model is also known as a toy
model for a porous medium used for exploration of low dimensional
percolation clusters \cite{baskin1}.

A special quantum behavior of a particle on the comb is the
quantum motion in the $d+1$ configuration space $(\mathbf{x},y)$,
such that the dynamics in the $d$ dimensional configuration space
$\mathbf{x}$ is possible only at $y=0$ and motions in the
$\mathbf{x}$ and $y$ directions commute. Therefore the quantum
dynamics is described by the following Schr\"odinger equation on a
comb
\be\label{qcom1}%
i\hh\frac{\prt\Psi}{\prt t}=\delta(y)\hat{H}(\mathbf{x})\Psi-
\frac{\hh^2}{2}\frac{\prt^2\Psi}{\prt y^2}\, , %
 \ee %
where the Hamiltonian is the same as in Eq. (\ref{fse1}), and
$\hH=\hat{H}(\mathbf{x})=\frac{\hh^2}{2}\nabla+V(\mathbf{x})$
governs the dynamics with a potential $V(\mathbf{x})$ in the
$\mathbf{x}$ space, while the $y$ coordinate corresponds to the 1d
free motion. All the parameters and variables are dimensionless
\cite{add2}. We will study an initial value problem with the
initial condition $\Psi(t=0)=\Psi_0(\mathbf{x},y)$.
It is worth noting that the semiclassical asymptotic expansion in
$\hh$ of the wave function is impossible, since the $\delta$
potential in the $y$ direction, ``cannot have a sensible
semiclassical limit'' \cite{Lary2,Lary3} and degrees of freedom
cannot easily be separated. Using the eigenvalue problem
 \be\label{qcom1_a}%
\hat{H}(\mathbf{x})\psi_{\lambda}(\bfx)=\lambda\psi_{\lambda}(\bfx)\,
,\ee %
we present the wave function in Eq. (\ref{qcom1}) as the expansion
$\Psi(\bfx,y,t)=\sum_{\lambda}\phi_{\lambda}(y,t)\psi_{\lambda}(\bfx)$,
where $\sum_{\lambda}$ also supposes integration on $\lambda$ for
the continuous spectrum. For the fixed $\lambda$ we arrive at the
dynamics of a particle in the $\delta$ potential
\be\label{qmay22a} %
 i\hh\frac{\prt\phi_{\lambda}}{\prt
t}=\frac{\hh^2}{2}\frac{\prt^2\phi_{\lambda}}{\prt
y^2}+\lambda\delta(y)\phi_{\lambda} \, . %
\ee %
The Green function for this Schr\"odinger equation has been
obtained in \cite{Lary1,Lary2}
\bea\label{qcom2_a}%
G_{\lambda}(y,t;y')=G_0(y,t;y')+ \nonumber \\
-\frac{\lambda}{\hh}\int_0^{\infty}
du G_0(|y|+|y'|+u,t:0)\exp(-u\lambda/\hh))\, , %
\eea %
where $ G_0(y,t;y')=\frac{1}{\sqrt{2\pi i\hh t}}
\exp\left(i\frac{(y-y')^2}{2\hh t}\right)$ is the free particle
propagator. Using this result, we obtain for the wave function of
Eq. (\ref{qcom1})
\bea\label{qcom2_b} %
&\Psi(\bfx,y,t)=\int dy'G_0(y,t;y')\Psi_0(\bfx,y') +\nonumber \\
&-\int
dy'duG_0(|y|+|y'|+u,t;0)e^{-u\hH/\hh}\frac{\hH}{\hh}\Psi_0(\bfx,y')\,
.\eea %

Now our aim is to compare Green's function of this solution with
the one of Eq. (\ref{fse3}). Contrary to fractional diffusion,
where the FFPE with $\alpha=1/2$ is identical to the comb model,
the FSE (\ref{fse1}) is not identical to the quantum comb model of
Eq. (\ref{qcom1}). Nevertheless, the models have some features in
common, namely there are the some singularities of the Green
functions which determine the spectrum. To show this, let us
present both Eq. (\ref{fse3}) and Eq. (\ref{qcom2_b}) in the form
of the inverse Laplace transform. For simplicity we consider the
one-dimensional $x$ space. First, we consider Eq. (\ref{qcom2_b}).
Presenting the wave function as the Laplace inversion and carrying
out integration on $u$ in the second part, we have for the Green
function
\bea\label{qcom2_c} %
&\hat{G}(x,y,t;x',y')=G_0(y,t,y')\delta(x-x')- \frac{1}{2\pi
i}\int_{\sigma-i\infty}^{\sigma+i\infty}\sum_{\lambda}
\nonumber \\
&\frac{\exp\Big(st +\sqrt{-2is/\hh}||y|+|y'||\Big) d s}{\sqrt{i\hh
s}\Big(\sqrt{-2is/\hh}-\lambda/\hh\Big)}\Psi_{\lambda}^*(x')\Psi_{\lambda}(x)\, , %
\eea %
where the eigenvalue problem of Eq. (\ref{qcom1_a}) is used. Using
integral presentation of the Mittag-Leffler function \cite{BE}, we
rewrite the Green function of the FSE in Eq. (\ref{fse3}) as
follows
\be\label{qcom6} %
G(x,t;x')=\sum_{\lambda}\frac{1}{2\pi i}
\int_{\sigma-i\infty}^{\sigma+i\infty}
\frac{\psi_{\lambda}^*(x')\psi_{\lambda}(x)e^{st}ds}{\sqrt{s}
(\sqrt{s}-\lambda/\sqrt{2i\hh})}\,
. \ee %
Therefore, disregarding the free propagator in Eq.
(\ref{qcom2_c}), one obtains the same spectral properties of both
systems, since the spectral decompositions of the evolution
operators are the same.

In what follows we consider the dynamics of the FSE in the
framework of the Green function (\ref{qcom6}). There are two
contributions to this integral from a pole at
$s_0=\lambda^2/2i\hh$ and from a branch point at $s=0$ with a
branch cut from $-\infty$. Carrying out this integration (see
\cite{naber}), we present the wave function in the following
operator form (see Eq. (\ref{fse3}))
\bea\label{qcom7} %
&\psi(x,t)=E_{\frac{1}{2}}
\left(\hH\left[\frac{t}{2i\hh}\right]^{\frac{1}{2}}\right)\psi_0(x)
\nonumber \\%
&=\left[2e^{\frac{-i\hH^2t}{2\hh}}-\frac{\hH\sqrt{2i\hh}}{\pi}\int_0^{\infty}
\frac{e^{-rt}dr}{\sqrt{r}(2i\hh r+\hH^2)}\right]\psi_0(x)\, . %
\eea  %
It consists of two parts, the oscillatory one and the decay in
time. This expression is convenient for studying the large time
asymptotic $t\gg 1$, when the decay term can be neglected.

The decay term is of the order of $\sim\sqrt{\hh}$, and can be
neglected in the semiclassical limit $\hh\rightarrow 0$ as well.
The evolution of the semiclassical wave function $\psi_{\rm
scl}(x,t)=\psi(x,t)/2$ is due to the oscillatory term with the
Hamiltonian ${\cal H}_{\rm scl}=\hH^2/2$. This semiclassical
dynamics is essentially nonlinear.

As an example of the Hamiltonian, we take the same operator
considered in Ref. \cite{bi2004} for L\'evy walks on the comb. It
is
\be\label{qex1}%
\hH=-2i\hh\omega \left(x\frac{\prt}{\prt x}+\frac{1}{2}\right)\, , %
\ee%
where $\omega$ is a dimensionless frequency.  In the framework of
the Fokker-Planck equation, it corresponds to the inhomogeneous
convection. Here this Hamiltonian describes a one-dimensional
motion near a hyperbolic point, and it has been studied in
connection with the Riemann zeros and eigenvalue asymptotics
\cite{berry1,berry2,armitage}, scattering of the inverted harmonic
oscillator \cite{bhaduri}, and eigenstates near a hyperbolic point
\cite{nonnemacher}. 

The standard quantum dynamics with the Hamiltonian of Eq.
(\ref{qex1}) is linear and coincides with semiclassical one. The
wave functions and expectation values are well behaved values. For
example, evolution of the wave function is an explicit function of
the initial condition
\be\label{scl1a} %
\Psi_{\hH}(x,t)=e^{-i\hH t/\hh}\psi_0(x)=
e^{-\frac{t}{2\hh}}\psi_0\left(xe^{-\frac{t}{\hh}}\right)\, ,
\ee%
and defining $y=xe^{-\frac{t}{\hh}}$, one obtains for the second
moment
\be\label{scl1b} %
\lgl \hat{x}^2(t)\rgl=e^{\frac{2t}{\hh}}\int
y^2\psi_0^{*}(y)\psi_0(y)dy < \infty\, . %
\ee%
Contrary to this, the semiclassical dynamics of the system
described by the FSE (and on the comb, respectively) is nonlinear
and the expectation values can diverge for the same initial
condition $\psi_0(x)$:
\be\label{scl2a} %
\psi_{\rm scl}(x,t)= \sqrt{\frac{\hh}{2\pi i
t}}\int_{-\infty}^{\infty}du
e^{\frac{i\hh u^2}{2t}-\frac{u}{2}}\psi_0\left(e^{-u}x\right)\, , \ee %
\be\label{scl2b} %
\lgl \hat{x}^2(t)\rgl=\int
y^2\psi_0^{*}\left(ye^{\frac{it}{\hh}}\right)\psi_0\left(ye^{\frac{-it}{\hh}}\right)dy\,
. \ee %
For example, for the Gaussian initial condition
$\psi_0(x)=e^{-x^2}/\sqrt{\pi}$ the expectation value in Eq.
(\ref{scl2b}) diverges at $t=\hh\pi/4$. This result was observed
in Ref. \cite{bervish} where the quantum motion near the
separatrix was studied.

We can conclude that the fractional time derivative, at least for
$\alpha=1/2$, reflects an effective interaction of a quantum
system with an additional degree of freedom. As already mentioned,
the $y$ direction in the comb model was introduced to model the
time fractional derivative $\frac{\prt^{\frac{1}{2}}}{\prt
t^{\frac{1}{2}}}$ by analogy with the continuous time random walk,
where delay times of escapes from the motion in the $\bfx$ space
are distributed by the power law $\sim t^{-3/2}$
\cite{em1,baskin1,bi2004}.  It is worth noting that in the
subdiffusive case with $\alpha=1/2$ the fractional Fokker-Planck
equation is identical to the diffusion comb model \cite{bi2004}.
In the quantum case the situation differs essentially from
fractional diffusion. First of all, the quantum comb model and FSE
are not identical. As shown here the systems are isospectral: they
have the same singularities for the Green functions. In the
quantum case a ``random entrapping'' is due to the reversible
leakage probability of the wave function. The decay term in the
Green function of Eq. (\ref{qcom7}) is the specific property of
the FSE which violates the Hermitian property of Hamiltonian
$\hH$. For the quantum comb, the $y$ space is the environment. In
this connection, an interesting question arises: how does the time
fractional derivative, which describes a specific interaction with
the environment, relate to a possible description in the framework
of the Lindblad equation \cite{lindblad,add3}? Both the quantum
motion on the comb and the FSE introduce new nonlinear phenomena
in the semiclassical limit, and this semiclassical approach
differs from one described in the framework of the standard
Schr\"odinger equation.

I am grateful to Prof. L. Schulman for having drawing my attention
to the result of Ref. \cite{Lary2,Lary1}. The hospitality of the
Max--Planck--Institute of Physics of Complex Systems (Dresden),
where a part of the work was performed, is gratefully
acknowledged. This research was supported by the Israel Science
Foundation.


\begin{thebibliography}{99}
\bibitem{add2} Analogously to the FSE (\ref{fse1}), following Ref.
\cite{naber}, one introduces the Planck length $L_P=\sqrt{\hbar
G/c^3}$, time $T_P=\sqrt{\hbar G/c^5}$, mass $M_P=\sqrt{\hbar
c/G}$, and energy $E_P=M_Pc^2$, where $\hbar,~G,$ and $c$ are the
Planck constant, the gravitational constant and the speed of
light, respectively. Therefore, quantum mechanics of a particle
with mass $m$ is described by the dimensionless units
$x/L_P\rightarrow x,~y/L_P\rightarrow y,~ t/T_P\rightarrow t$,
while the dimensionless Planck constant is defined as the inverse
dimensionless mass $\hh=M_P/m$. Note, that the dimensionless
potential is now $V(\bfx)\rightarrow V(\bfx)/M_Pc^2$.
\end{thebibliography}
\end{document}